\documentstyle[preprint,aps]{revtex}

\begin{document}
\voffset=.75truein
\hoffset=.75truein

\draft\title{Evolution of the Potential in Cosmological Gravitational
Clustering}

\author{Adrian L. Melott and Jennifer L. Pauls}
\address{Department of Physics and Astronomy, University of
Kansas, Lawrence, Kansas 66045}
\date{\today}
\maketitle
\begin{abstract}
In recent years there has been a developing realization that the interesting
large--scale structure of voids, ``pancakes", and filaments in the Universe is
a consequence of the efficacy of an approximation scheme for cosmological
gravitation clustering proposal by Zel'dovich
in 1970.  However, this scheme was only supposed to apply to smoothed initial
conditions.

We show that this can be explained by the fact that the gravitationally evolved
potential from N--body simulations closely resembles the smoothed potential of
the initial conditions.  The resulting ``hierarchical pancaking" picture
effectively
combines features of the former Soviet and Western theoretical pictures for
galaxy and large--scale structure formation.
\end{abstract}
\vskip .25in
\pacs{PACS Numbers:  98.80 Dr, 95.30 Sf}
\newpage
Over the last forty years, it has been realized that galaxies are clustered in
space, which is measurable by a two--point correlation function\cite{pjep80}.
During
the last twenty years, it has become clear that the distribution has a rich,
frothy structure of voids, filaments, and sheets\cite{pjep93}. Gravitational
instability is the dominant theory for explaining the development of clumpy
structure in the Hot Big Bang Cosmology.  Yet there is a problem in accounting
for all the coherent anisotropic structure.

In 1970 Zel'dovich\cite{yabz} proposed an approximation for clustering which is
essentially inertial motion in comoving coordinates.  Specialized to critical
density models, it is
$$\bar x(\bar q,t)=\bar q+ a(t)\bigtriangledown S(\bar q)\eqno (1)$$
where $\bar x$ is the (Eulerian) position, $\bar q$ is the initial
(Lagrangian) coordinate, $a$ is the scale factor of the Universe, and $S$ is a
velocity potential related by a constant factor to the gravitational potential
$\phi$.  Divergence in $\phi$ is prevented by of doing the convention change
of
variables to density contrast $\rho - {\overline{\rho}} \over
{\overline{\rho}}$ as the source term for the
Poisson equation.  The pseudo--Newtonian treatment of cosmological
gravitational clustering has recently been put
on a firm footing\cite{{lak},{ebah}}

In spite of the general success of the gravitational instability picture, there
is no ready explanation for the coherence of large scale structure, i.e
superclusters.  The Zel'dovich approximation (ZA) does predict such
structures\cite{sfh},
but ZA depends on the assumption of long--range coherence of the
gravitational potential (otherwise particles should be deflected from their
trajectories by very small--scale inhomogeneities).  Theories with smoothed
initial conditions,
such as ``Hot Dark Matter", apparently have trouble making any structures in
voids, which results in an excessive correlation amplitude\cite{sdmw}.

Hierarchical clustering models, which have initial density fluctuations on all
scales, are more generally successful.  Such models are typically specified by
the power spectrum of density fluctuations
$$P (k) \equiv < \delta^2_k>\eqno (2)$$
where the $\delta_k$ are Fourier components.  Power laws
$$P(k)\propto k^n\eqno (3)$$
are useful for theoretical analysis and discussion.  The most favored models
today have $n\sim 1$ for
small $k$, possibly a relic of inflation, and turn over gradually to negative
$n$
at large $k$, depending on the matter contents of the Universe.  Such models
usually are based on collisionless dark matter.

In spite of the fact that these theories do not have smoothed initial
conditions, evidence began accumulating from numerical simulations that they
produced interesting large--scale anisotropic structures\cite{{alm},{md}}. As
this
became an accepted feature of such models, an explanation was developed based
on the adhesion approximation\cite{{sng},{lket}}. This explanation contains the
correlation length of $\phi$
as a crucial feature, and consequently identifies
$n =-1$
as a transitional power law.

However, this argument must be incomplete.  Use of ZA can predict structure
very well if the initial conditions are smoothed, removing small-scale initial
power.  The best smoothing appears to be Gaussian convolution around the scale
of nonlinearity\cite{alm94} More recently, a detailed study of the behavior of
the
approximation (which we call TZA for truncation of the spectrum, to distinguish
it from ZA) over a range $-3 \le n \le +3$ has been made with surprising
results\cite{jlp}.
The performance of TZA degrades as $n$ increases but is still quite good
even for the extreme case $n = +3$; it is far better than conventional Eulerian
linear perturbation theory for example.  One can use the initial conditions to
predict the location and orientation of filament--like objects with
considerable accuracy.  N--body simulations appear to be no longer necessary
for most spectra of interest if one is satisfied with resolving galaxy group
mass scales [12].

An explanation is now possible for the unreasonable utility of TZA; more detail
is presented elsewhere [13].  Because the growth of fluctuation amplitude
compenstates for expansion,
the peculiar gravitational potential
evolves to linear order such that $\phi$
is
constant. We discuss for changes beyond linear order.

In Figure 1 we show the initial and nonlinear evolved potential along a
diagonal of each of four simulation cubes.  We did power--law N--body
simulations with $n$ = --3, --1, +1, and +3.  These are 128$^3$ PM
simulations\cite{amss}.
The moment
chosen is that when the scale of nonlinearity has grown to $k_{n\ell} = 8k_f$,
where
$k_f$ is the fundamental mode of the cube.

For $n$ = --3 and --1 the evolved potential is very close to the initial, as
expected from trivial considerations (the power spectrum of the potential is $n
-4$, so long waves dominate).  For $n = +1$ and $+3$ there is no real
resemblance between the potentials,
although the eye can detect some correlation for $n = +1$.

Computation of the coherence length of the potential shows that it is largely
unchanged by evolution for $n = -3$ and $-1$.  However it grows by a factor of
more than 7 for $n = +1$ and $+3$.  (In both cases the number should be larger;
we
are resolution limited, especially for +3).  Adhesion arguments are based on
coherence of the initial potential, and so missed making the prediction in the
latter two cases.

Now let us compare the evolved potential to the smoothed initial potential.
This is shown in Figure 2 with a scale change for clarity.  There is a strong
resemblance, decreasing with $n$.  It is clear from this that modestly
nonlinear evolution (up to $\delta_\rho/\rho\sim 1$ from linear extrapolation)
can be accurately
described by constant gravitational potential with increasing smoothing.  We
have crosscorrelated the initial with the final potential.  There is a
significant
signal, increasing predictably as $n$ decreases.  Modes $k <k_{n\ell}$ are
known to
grow linearly.  However, when we smooth the initial potential by Gaussian
convolution, there is an enormous increase in signal, verifying the visual
impression of comparing Figures 1 and 2.  The smoothing windows found to work
best for TZA [12]
were used for this comparison.  In the Table we show further information.
More details on this and other detailed
comparisons between the simulations at TZA are given elsewhere [13].

The very strong resemblance of the nonlinear potential to the smoothed initial
potential explains why TZA works so well over such a wide range of spectral
indices.
The clumps which have formed by hierarchical clustering are moving in a
background potential which is close to a smoothed version of the initial.
Also, adhesion and other approximations are limited\cite{almapj} by their use
of the
initial potential.  Even for $n$ = --1 we find some improvement by smoothing.
This suggests that a new class of second--order approximation schemes can be
constructed which go beyond ZA but use the smoothed potential.

The Universe possesses a rich large--scale structure because gravitational
clustering smooths the potential.  This is a somewhat counter--intuitive
result, but there have been precursor hints, for example based on the topology
of large--scale structure\cite{almet}. Furthermore, the smoothed and linearly
evolved
density field
manifestly does not resemble the nonlinear density field\cite{pc}.
Hierarchical
clustering (largely developed in the West) is a good description of
small--scale clumping and gives reasonable results for galaxy formation [2].
However, the motion of the clumps is driven by the smoothed potential, which
brings into play all the machinery developed by the ``Moscow school" during the
70's and 80's [6].  It appears that a unified picture of galaxy and
large--scale structure formation can now emerge.

\section{Acknowledgements}

We thank NSF (AST--9021414), NASA (NAGW--3832), and the National Center for
Supercomputing Applications for support.  Extremely useful conversations were
those with Jim Peebles, Sergei Shandarin, Lev Kofman, Dima Pogosyan, David
Weinberg, Bob Scherrer, and other participants in the 1994 Aspen Center for
Physics summer workshop on Gravitational Clustering in Cosmology.

\newpage
\centerline{FIGURE CAPTIONS}

Fig. 1. The peculiar gravitational potential is shown along one diagonal of
each of the four simulation cubes for (a) $n=-3$ (b) $n=-1$ (c) $n=+1$ (d)
$n=+3$. The units are arbitrary but do reflect the fact that the potential is
constant to linear order. The dotted line is the initial potential is constant
to linear order. The dotted line is the initial potential and the solid line
the evolved. Note strong evolution for $n\geq 1$.

Fig. 2. The potential is plotted in the same way as Figure 1 except that the
intial potential is smoothed by Gaussian convolution, and the vertical scale
has been expanded for $n=+1,+3$.
\newpage
\begin {table}
\centering
\caption{Cross-correlation of gravitational potentials.}
\label{symbols}
\begin{tabular}{cccc}
Spectral Index & Smoothed Initial/Initial & Initial/Final &
Smoothed Initial/Final \\
-3 & -- & 0.96 & -- \\
-1 & 0.999 & 0.987 & 0.990 \\
+1 & 0.68 & 0.65 & 0.94 \\
+3 & 0.14 & 0.10 & 0.69 \\
\end{tabular}
\end{table}

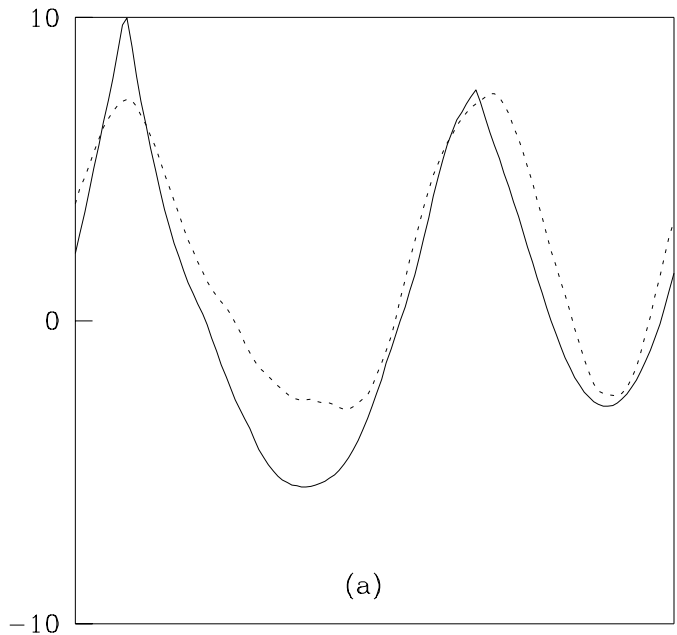
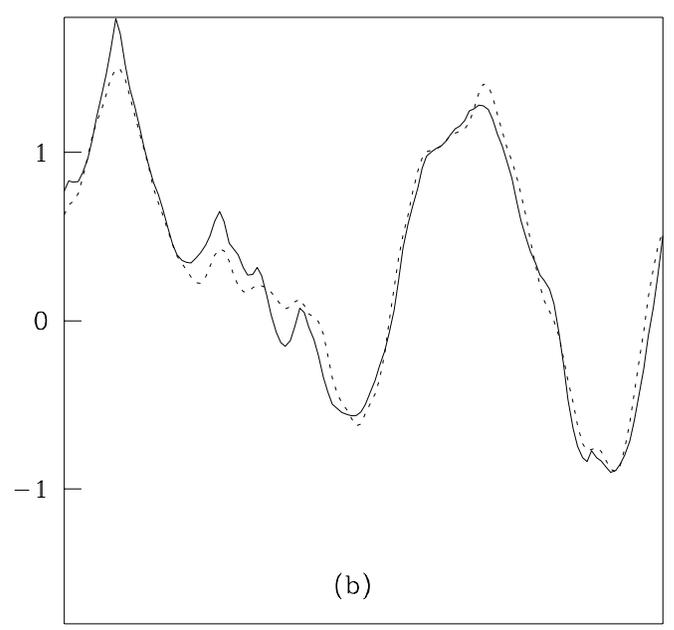
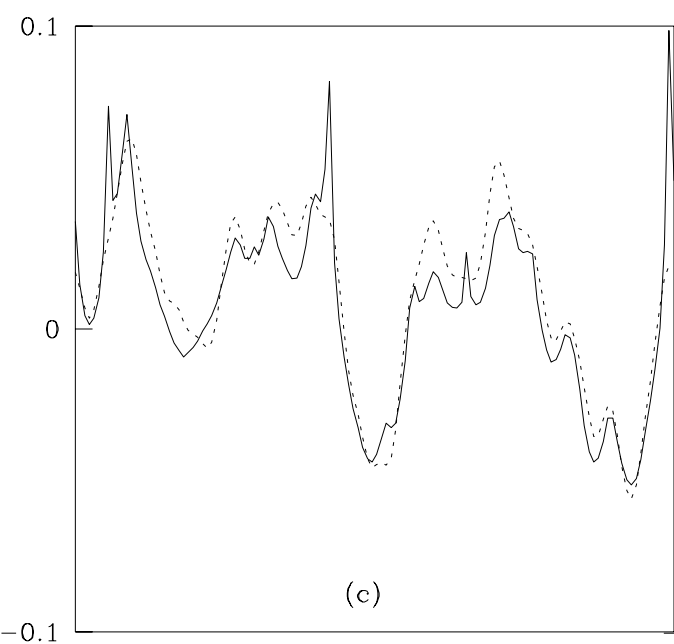
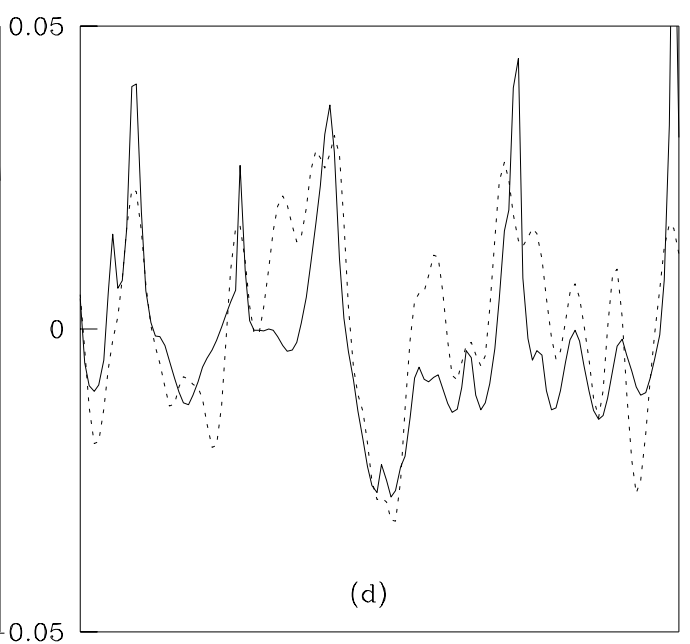
\end{document}